\documentclass[conference]{IEEEtran}
\IEEEoverridecommandlockouts
\usepackage{comment}
\usepackage{booktabs}
\usepackage[export]{adjustbox}
\usepackage{booktabs}
\usepackage{float}
\usepackage{flushend}
\usepackage{graphicx}
\usepackage{ifpdf}
\usepackage[utf8]{inputenc}
\usepackage{listings}
\usepackage{multirow}
\usepackage[caption=false,font=footnotesize]{subfig}
\usepackage{url}
\usepackage{tabularx}
\usepackage{textcomp}
\usepackage{xcolor}
\usepackage{xspace}
\usepackage[T1]{fontenc}
\usepackage{lmodern}
\usepackage{lipsum}

\definecolor{listinggray}{gray}{0.95}
\definecolor{darkgray}{gray}{0.7}
\definecolor{commentgreen}{rgb}{0, 0.4, 0}
\definecolor{darkblue}{rgb}{0, 0, 0.6}
\definecolor{purple}{rgb}{0.6, 0, 0.6}
\definecolor{middleblue}{rgb}{0, 0, 0.75}
\definecolor{darkred}{rgb}{0.4, 0, 0}
\definecolor{brown}{rgb}{0.5, 0.5, 0}
\definecolor{dkgreen}{rgb}{0,0.5,0}
\definecolor{orange}{rgb}{1,.5,0}
\definecolor{dandelion}{cmyk}{0,0.29,0.84,0}

\usepackage[normalem]{ulem}
\makeatletter
\def\cyanuwave{\bgroup \markoverwith{\lower3.5\p@\hbox{\sixly \textcolor{cyan}{\char58}}}\ULon}
\def\reduwave{\bgroup \markoverwith{\lower3.5\p@\hbox{\sixly \textcolor{red}{\char58}}}\ULon}
\def\blueuwave{\bgroup \markoverwith{\lower3.5\p@\hbox{\sixly \textcolor{blue}{\char58}}}\ULon}
\font\sixly=lasy6 
\makeatother

\def\BibTeX{{\rm B\kern-.05em{\sc i\kern-.025em b}\kern-.08em
    T\kern-.1667em\lower.7ex\hbox{E}\kern-.125emX}}

\newif\ifdraft{}

\ifdraft{}
  \newcommand{\aanote}[1]{ \textcolor{blue} { ***aymen: #1 }}
  \newcommand{\jhanote}[1]{ {\textcolor{red} { ***shantenu: #1 }}}
  \newcommand{\mtnote}[1]{ {\textcolor{orange} { ***matteo: #1 }}}
  \newcommand{\kyle}[1]{ {\textcolor{purple} { ***kyle: #1 }}}
  \newcommand{\logan}[1]{ {\textcolor{green} { ***logan: #1 }}}
  \newcommand{\ian}[1]{ {\textcolor{violet} { ***ian: #1 }}}

\else
  \newcommand{\aanote}[1]{}
  \newcommand{\jhanote}[1]{}
  \newcommand{\mtnote}[1]{}
  \newcommand{\kyle}[1]{}
  \newcommand{\logan}[1]{}
  \newcommand{\ian}[1]{}
\fi

\newcommand{\UP}{\vspace*{-1.0em}}
\newcommand{\up}{\vspace*{-0.5em}}

\lstdefinestyle{myListing}{
  frame=single,
  backgroundcolor=\color{listinggray},
  language=C,
  basicstyle=\ttfamily \footnotesize,
  breakautoindent=true,
  breaklines=true
  tabsize=2,
  captionpos=b,
  aboveskip=0em,
  belowskip=-2em,
}

\lstdefinestyle{myPythonListing}{
  frame=single,
  backgroundcolor=\color{listinggray},
  language=Python,
  basicstyle=\ttfamily \footnotesize,
  breakautoindent=true,
  breaklines=true
  tabsize=2,
  captionpos=b,
}

\ifpdf{}
  \DeclareGraphicsExtensions{.pdf, .jpg, .tif}
\else
  \DeclareGraphicsExtensions{.eps, .jpg, .ps}
\fi

\newcommand*\rot{\rotatebox{90}}

\begin{document}

\bstctlcite{IEEEexample:BSTcontrol}

\title{RADICAL-Pilot and Parsl: Executing Heterogeneous Workflows on
HPC Platforms
}


\author{Aymen Alsaadi$^{1}$, Logan Ward$^{2}$, Andre Merzky$^{1}$, Kyle Chard$^{2}$$^{,4}$, Ian Foster$^{2}$$^{,4}$, Shantenu Jha$^{1}$$^{,3}$, Matteo Turilli$^{1}$$^{,3}$\\
   \small{\emph{$^{1}$Rutgers, the State University of New Jersey, Piscataway, NJ 08854, USA}}\\
   \small{\emph{$^{2}$Data Science and Learning Division, Argonne National Laboratory, Lemont, IL 60439, USA}}\\
   \small{\emph{$^{3}$Brookhaven National Laboratory, Upton, NY 11973, USA}} \\
   \small{\emph{$^{4}$Department of Computer Science, University of Chicago, Chicago, IL, USA}} \\
}

\maketitle

\begin{abstract}
Workflows applications are becoming increasingly important to support scientific
discovery. That is leading to a proliferation of workflow management systems
and, thus, to a fragmented software ecosystem. Integration among existing
workflow tools can improve development efficiency and, ultimately,
increase the sustainability of scientific workflow software.
We describe our experience with integrating RADICAL-Pilot (RP) and Parsl as a
way to enable users to develop and execute workflow applications with
heterogeneous tasks on heterogeneous high performance computing resources. We
describe our approach to the integration of the two systems and detail the
development of RPEX, a Parsl executor which uses RP as its workload manager. We
develop a RP executor that executes heterogeneous MPI Python functions
on CPU cores and GPUs. We measure the weak and strong scaling of RPEX, RP
and Parsl when providing new capabilities to two paradigmatic use cases: Colmena
and Ice Wedge Polygons.
\end{abstract}

\begin{IEEEkeywords}
Workflows, HPC, MPI executor, middleware integration.
\end{IEEEkeywords}

\section{Introduction}
\label{sec:intro}

\ian{The following comments may or may not be helpful; do with them what you
will:  \begin{quotation} I find it quite hard to tell what is being done in the
paper, in my view because the lengthy philosophical discussions about the
general value of combining things means that the really neat details of what
specifically is being done gets lost. I wonder if some rephrasing and
restructuring could help. As I understand things, the basic idea here is simple
and smart --- \emph{use RP as a new executor for Parsl, in order to support many
MPI computations more efficiently than with other Parsl executors}. If so,  then
the paper could be structured as: 1) why we want to support many MPI executors,
and why current Parsl executors don't do that well; 2) how "RPEX" is constructed
[using that name to be consistent with other Parsl executors], 3) application
studies, 4) discussion of the broader lessons about value of integration. in my
view, make the whole paper would be much clearer.  The title of the paper could
be changed, to e.g., ``Massively scalable Python scripting: A Radical Pilot
executor for Parsl" or ``RPEX: Leveraging Radical Pilot to Enhance Parsl
Scalability", or similar. Probably some technical details, e.g. about how RP
works internally, could be scaled back.\end{quotation} }

\kyle{I've only read the intro.. but I agree that the messaging could be
streamlined. The intro currently talks about exascale (which seems not too
important), different task types, integration of middleware components. I think
it would be useful to pick one angle (e.g., enhancing Parsl to support MPI tasks
important for a range of applications, or interoperability between workflow
systems. I like Ian's suggested structure, and the RPEX name.}

\jhanote{Chiming in a bit late, so timing may make the limited ideas even less
useful. I think the two approaches Kyle outlines are not mutually exclusive.
Both angles are possible, maybe not co-equally, but should be leveraged. This
paper was conceived as an ExaWorks paper (and clearly has the fingerprints of
its original conception). As conceived it was meant to be an ``experience in
integration paper. But it is clearly both an experience paper, but one that
establishes the advantages of integration between otherwise distinct tools used
to support scientific/HPC workflows. Thus, my two cents would be to retain the
former angle -- while reducing and sharpening the writing -- and to also
highlight the fact that this work is motivated by well defined need, viz.,
support many MPI executors for ParSL, and delivery of something (possibly)
production grade, viz., RPEX . That said, one possible approach could be a
variant on Ian's proposed structure: (1) Importance and need of integrating
tools/software to manage proliferation (e.g., 1001 workflow systems) and
targeted performance-engineering (optimized components); (2) Need for many MPI
executors; (3) RPEX ; (4) Application Case studies; and (5)
Conclusions/discussion}

\kyle{That makes sense to me. I think selling it as a case study of integration
is good, and highlighting experiences. E.g., by adopting a standard interface
things are easier, but we still have to deal with different task representations,
result, running different types of tasks, etc.}

\mtnote{I think the integrative element of this paper is important. In the
context of Exaworks, our two groups are championing an approach that breaks away
from a fairly established model in which each group develops an end-to-end
stack, replicating many capabilities that are already available in other stacks.
I would like for our paper to make this point clearly and explicitly as I
believe it can benefit our community as a whole. Otherwise, I think we are
incorporating most of the other suggestions, finding a balance between
sharpening the scope of the paper, and avoiding major rewriting with little time
left before the deadline and a very low page limit. Please let me know whether
you think we managed to reach a viable introduction and, once all the comments
are addressed, a viable draft as a whole.}



\kyle{How about we start with something like the following.. Workflow systems
are becoming increasingly ubiquitous as they effectively abstract the complexity
of orchestrating the execution of many heterogeneous tasks across diverse
computing resources. The broad adoption of workflows has led to the development
of hundreds of distinct workflow systems (e.g., one list maintained by the CWL
community lists 318 workflows systems~\cite{workflow-systems-url}); however,
there is significant overlap between the goals and capabilities provided by many
of these workflow systems, making it difficult for users to evaluate and select
the most suitable workflow system for a given application. Further, the
development of these workflow systems is inefficient as there is significant
duplication of functionality between workflow systems and a lack of robustness
as it is infeasible for a single workflow system to meet all application
requirements on all potential resources. A recent series of community summits
organized by the workflows community~\cite{} highlighted the need for workflow
system interoperability as a way of reducing development inefficiency; improving
robustness, performance, and portability; and ultimately enhancing the
sustainability of the workflows community. We present here our experiences
integrating two workflow systems to support use cases that cannot be met by each
workflow system individually. }

\jhanote{Matteo, Kyle: Your suggestions and considerations make sense to me.
Thanks. Not sure if either of you are working on the introduction but glad to
take a crack at refining on Monday AM as needed}

Workflow systems are becoming ubiquitous as they effectively abstract the
complexity of orchestrating the execution of heterogeneous tasks across diverse
computing resources. This has led to the development of hundreds of workflow
systems~\cite{workflow-systems-url} with significant overlap between their goals
and capabilities.
The development of these systems is inefficient: there is significant
duplication of functionality and a lack of robustness as it is infeasible for a
single workflow system to meet all application requirements on all potential
resources. Recent summits organized by the workflows
community~\cite{workflows2021silva-1} highlighted the need for workflow system
interoperability as a way of reducing development inefficiency; improving
robustness, performance, and portability; and ultimately enhancing the
sustainability of the workflows community.




Task heterogeneity is fundamental to high performance computing (HPC) scientific
scientific workflows. Tasks may be standalone executables or functions
implemented in diverse programming languages. Both executable and function tasks
have diverse requirements: from single-core utility functions to multi-node MPI
simulation executables. Supporting such task heterogeneity requires: (1) a
workflow system for users to express and execute applications with diverse types
of tasks; and (2) a workload manager capable of interpreting and managing the
execution of those tasks at scale and on diverse HPC platforms.

Here, we present the integration of the Parsl~\cite{babuji2019parsl} workflow
system and the RADICAL-Pilot (RP)~\cite{merzky2021design} workload manager,
independently developed by different research groups. We adopt a loosely-coupled
integration approach, developing a RP Executor (RPEX) for Parsl and a new RP
executor to add the capability to distribute and execute MPI Python functions
concurrently to (non)MPI executable tasks.
Our integration brings new capabilities to both systems: Parsl users can benefit
from the scalable and performant RP runtime capabilities with minimal or no
changes to their code, while RP users gain a larger choice when deciding what
workflow system to use for their applications, e.g., Parsl, EnTK, Swift.

We describe how our integration brings new capabilities to two use cases.
Colmena~\cite{ward_colmena_2021}, a Python package that uses Parsl to execute
ensemble applications, gains new MPI capabilities via RPEX, RP and its new
executor. Ice Wedge Polygons (IWP) benefits from Parsl's dataflow capabilities
and Python API to implement a workflow application that uses RPEX and RP to
concurrently execute multi-node MPI Python functions on CPUs and GPUs.

We measure strong and weak scaling of the new RP executor, and of RPEX for both
Colmena and IWP, showing that overheads are small or invariant of scale. We
compare the resource utilization of Colmena with RPEX, showing that is analogous
to that obtained in Ref.~\cite{ward_colmena_2021} with Parsl's HTEX executor.
The insight gained by our analysis shows the viability of the proposed
integrative approach and offers useful information about how to improve the
execution of heterogeneous tasks on HPC resources at scale.

\section{Related Work}
\label{sec:related}

The integration of workflow and runtime systems, and the building blocks
approach to workflow middleware~\cite{turilli2019middleware} extend
functionalities and interfaces, enabling different programming paradigms and the
execution of different applications on different platforms at a variety of
scales.

There have been multiple approaches to integrating the traditional big-data
middleware stack with HPC workflow and resource management
tools~\cite{wang2009kepelr_hadoop,Luckow_2016_Hadoop}. Here the primary focus
is on integration of traditional HPC software systems, for example, the
integration of PyCOMPS~\cite{badia2022pycompss} with other frameworks, or
Swift~\cite{wilde2011swift} with RP. In the later, the integration was based
on their application programming interfaces (API) and special-purpose
connectors~\cite{turilli2017evaluating}. This reduced the engineering effort
spent on each system, centering the development on a small and independent
component that translates computational requirements between the workflow
application layer, and the resource management and task execution layers.

\jhanote{I think the following two paragraphs can be reduced. Proposed rewrite above
for your consideration}
Shaffer et al.~\cite{shaffer2021lightweight} use the Parsl API in their
integration of Parsl with the WorkQueue framework~\cite{bui2011work}.
They achieve lightweight function monitoring across
HPC resources by
forking a new process for every executing function.
However, this approach introduces: (i) additional resource requirements due to
the creation of a monitoring process for each Python function executed, and (ii)
additional overhead associated with launching that extra process.

Merlin~\cite{peterson2019merlin} is designed to enable the execution of large
ensembles simulations and machine learning analyses on HPC platforms. Merlin
uses the Maestro~\cite{Maestro_2019} interface to define workflows of millions
of tasks, and deploys Flux~\cite{Dong_2014_Flux} to scale such workflows on HPC
systems. However, its use of Maestro's YAML-based interface to define tasks
restricts it to a shell syntax, making it challenging for the end-user to take
advantage of clearer and more powerful programming languages, such as Python.

\section{Use Cases}
\label{sec:usecases}

We present two exemplar use cases that require the capabilities of both RP and
Parsl to execute (non)MPI executables and Python function tasks on HPC
platforms.


\subsection{Colmena: Intelligent Steering of Ensemble Simulations}


\ian{I reworded the following to emphasize that the critical need for this
application is a new Parsl executor that can run many MPI computations at
once.}\logan{Thanks! It is much more to-the-point}

Colmena~\cite{ward_colmena_2021} is a Python package for machine learning-based
steering of ensemble computations on HPC platforms, for such purposes as fitting
interatomic potentials~\cite{guo2022composition}. A Colmena application is
organized as a \emph{Thinker} process that implements a strategy for selecting
computations that are then submitted to a \emph{Task Server} process for
execution.
Currently, Colmena Task Server uses the Parsl workflow engine to dispatch tasks
to multiple processors.

The computations managed by Colmena applications are frequently MPI programs of
modest scale.
Thus, in order to make efficient use of large parallel computers, Colmena needs
to run efficiently many MPI applications at once---something that existing Parsl
executors are not able to do. Parsl deploys MPI tasks using a single worker on
an HPC launch node
that is responsible for pre- and post-processing tasks and invoking the MPI
launcher using subprocesses. Large ensemble can lead to overheads of minutes, as
processing tasks compete for resources and requests overwhelm the MPI launcher.
Colmena is thus an excellent use case for the new RPEX executor.

\subsection{Ice Wedge Polygons}

Ice wedges are common permafrost subsurface attributes that evolved by
accumulated frost cracking and ice-vein growth over long periods of time. These
wedge-shaped ice masses create polygonized land surface patterns called Ice
Wedge Polygons (IWP) across large Arctic areas. Observing IWP requires
processing very high spatial resolution (VHSR) satellite imagery at multiple
spatial scales~\cite{Witharana_2021_IWP}.

IWP is implemented via MPI Python functions that require the concurrent use of
both GPUs and CPUs. Each image is processed by performing two
operations---tiling and inference.
Tiling uses CPUs to divide each image into
360$\times$360 pixels tiles; inference uses a GPU to extract the surface
patterns from each tile.

The RPEX executor offers the required workload runtime capabilities via
RADICAL-Pilot and a flexible programming model via Parsl to execute IWP
multi-node MPI Python functions concurrently on CPUs and GPUs.

\section{RADICAL-Pilot And Parsl Integration}
\label{sec:integration}

We integrate RP and Parsl into a system that we name, for simplicity, RPEX.
Importantly, we implement and extend an existing
interface \kyle{Do you develop one or implement a existing interface, i.e.,
Python executor interface?}\aanote{thanks, I tried to fix it. Is this ok @kyle
@mturilli}between the two systems `as they are,' providing users with the sum
of the two systems' capabilities, without engineering a whole new
system. \kyle{Somewhere, probably use cases, it would be good to say why we even
want the  Parsl interface. Is it that there are many apps already written with
Parsl and  they wont need to change? Is there some particular value to RP to
have the Parsl model on top? Maybe we could say that RP has focused on building
a flexible and fast runtime system and part of the benefit there is that it can
be easily adapted to different use cases via different front end interfaces
(e.g., Parsl for Python DAGs, EnTK for ensembles, etc.)}\aanote{@kyle, something like below?}
This integration allows RP to benefit from Parsl flexible programming model and
its workflow management capabilities to build dynamic workflows. Additionally,
RPEX will benefit Parsl by offering the heterogeneous runtime capabilities of RP
to support many MPI computations more efficiently than with other Parsl executors.

\subsection{RADICAL-Pilot (RP)}

RP is a scalable, modular, and interoperable pilot system, coded in Python, that
enables the execution of
heterogeneous workloads on heterogeneous HPC resources.
RP has four main components~\cite{merzky2021design}:
The \emph{Pilot Manager} and \emph{Task Manager}, which are executed on a user
resource or on the login node of an HPC platform; the \emph{Agent}, which is
executed on the compute nodes of the target HPC platform; and a \emph{MongoDB}
database, which is hosted on resources accessible via network by the other
components.
\aanote{based on Shantenu's suggestion I removed RP's arch. figure and replaced
it with a citation}


RP provides methods for efficiently and effectively scheduling, placing, and
launching independent tasks across multiple nodes. RP uses the pilot
abstraction~\cite{Turilli_2018_Comprehensive} to support the concurrent
execution of up to $10^6$ tasks on $10^3$ compute nodes with low
overheads~\cite{merzky2021design}.



Different from other pilot systems, RP supports tasks that may vary
simultaneously along four dimensions; (1) programming model, including
single/multi cores/GPUs with MPI, OpenMP, and single/multi-[threaded$|$process]
tasks; (2) scale, from 1 to 27,000 GPUs and/or 1 to 467,000 CPU cores; (3) task
duration, from >1 second to 24 hours or more; and (4) task packaging method:
both standalone executables and Python functions.

\subsection{The Parsl parallel Python programming library}

Parsl~\cite{babuji2019parsl} is a Python module that enables parallel execution
of Python functions and orchestration of functions into dataflow workflows.
Parsl users
decorate Python functions
to indicate opportunities for concurrent execution. Parsl relies on
\emph{futures} to abstract asynchronous execution: invocation of a Parsl app
returns a future to the calling program. The future's state is set only when the
app completes execution; an attempt to read the future before that time causes
the application to block. Parsl enables dataflow semantics by allowing
developers to pass futures between apps.

\begin{figure}[h]
    \centering
    \includegraphics[width=0.50\textwidth]{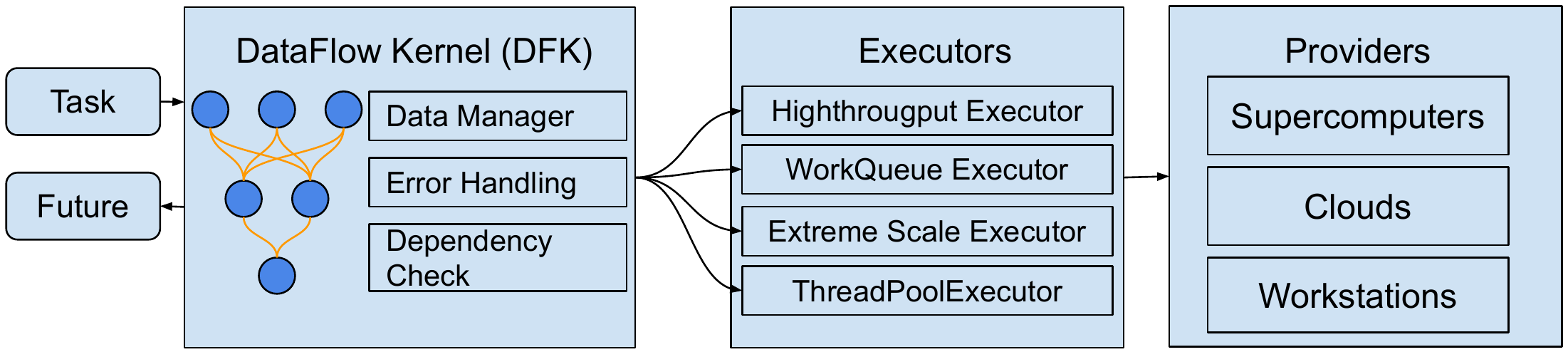}
    \caption{Parsl architecture and execution model. \jhanote{Is it possible
    to get a higher resolution version? Or is it just my old eyes ... ??}}\aanote{better?}
    \label{fig:parsl-arch-exec}
    \up
\end{figure}

Fig.~\ref{fig:parsl-arch-exec} shows the three main components of the Parsl
implementation: the Data Flow Kernel (DFK), Executor, and Provider.
Parsl applications start when a Python program calls a Parsl app and passes
either input arguments or a future from another app. The DFK wraps each task
with a Python future object~\cite{concc_2020}. Throughout execution, the DFK
maintains a directed acyclic graph (DAG) with nodes representing each invocation
of an app (called a `task') and edges representing futures passed between apps.
Once a task's dependencies are resolved, the DFK submits it to one or more
user-specified executors. The DFK tracks every task's state, updating the task
graph.

Parsl relies on Python's standard \texttt{concurrent.futures} executor interface
to dispatch tasks for execution. Parsl includes two in-built executors and an
external executor that implement this interface, each designed for a specific
type of workload. The high-throughput executor (HTEX) is a pilot-based executor
for rapid execution of many tasks.
The Extreme-Scale Executor (EXEX) executes tasks on a pool of multi-node
processes, using the Python package \texttt{mpi4py}~\cite{Dalcn_2005_mpi4py} to
build and manage communications between managers and workers. The WorkQueue
Executor (WQEX) uses WorkQueue to provide managed task execution with dynamic
resource sizing.

\subsection{Design}

RP's architecture~\cite{merzky2021design} and Parsl's architecture
(Fig.~\ref{fig:parsl-arch-exec}) suggest two integration points: using RP to
submit tasks to an existing Parsl executor or using the RP Agent as a new Parsl
executor. While the former integration point would extend Parsl's HPC resource
acquisition capabilities, it would not allow users to benefit from most of RP's
Agent capabilities. The latter integration point allows us to maintain Parsl API
and workflow-related capabilities while benefiting also of RP runtime
capabilities.


RP can provide MPI support across multiple HPC platforms, supporting multiple
dimensions of task and resource heterogeneity. Further, RP supports Single
Program Multiple Data (SPMD) and its performance is tailored to extreme-scale on
HPC resources~\cite{merzky2021design}. It also supports the concurrent execution
of multiple pilots on multiple HPC platforms~\cite{turilli2016integrating} and
the scheduling of a single workload across those
pilots~\cite{turilli2017evaluating}.

Integrating RP as a Parsl executor requires aligning the two systems' task
execution models.
%
RP's tasks are fully-decoupled, i.e., they have no data dependencies or those
dependencies have been already satisfied out-of-band. Each task is assumed to be
self-contained, executed by RP as a black-box that either returns or fails. RP
has no
knowledge of the code each task executes, enabling a separation of concern
between resource and execution management, and task executables. Consistently,
at application level, RP implements a `batch-like' programming model in which
groups of tasks (i.e., workloads) are described and submitted for execution.
Concurrency is implicit:
once submitted, RP executes tasks with the maximum concurrency allowed by the
available resources.



Unlike RP tasks, Parsl tasks have dynamic data dependencies that must be
respected before execution.
At the application level, Parsl allows for the expression of nested parallelism
within a single task or across a batch of tasks. Parsl programming model enables
various parallel computing paradigms such as procedural and dynamic workflow
execution and interactive parallel programming.

\kyle{This is the key info I think}
Parsl's tasks are Python functions while RP tasks are Python dictionaries
that are dynamically updated to reflect the state of the tasks. The difference
in the task object's type
is a communication barrier
that we overcame
by implementing a mid-point component called ``Task Translator'', with the
following capabilities: (i) detect whether Parsl task is a pure Python function
or a Python call to a Bash command; (ii) translate Parsl tasks into RP tasks;
and (iii) update the status of Parsl tasks, according to callbacks from RP
tasks.

Fig.~\ref{fig:int-arch} illustrates the translation of Parsl tasks into RP
tasks. Each Parsl task is translated via a direct (1:1) mapping in accordance to
the task submission criteria of Parsl's DFK. Thus, tasks are created at
application level and submitted to the executor one by one, iteratively.

\subsection{Implementation}

\begin{figure}[h]
    \centering
    \includegraphics[width=0.33\textwidth]{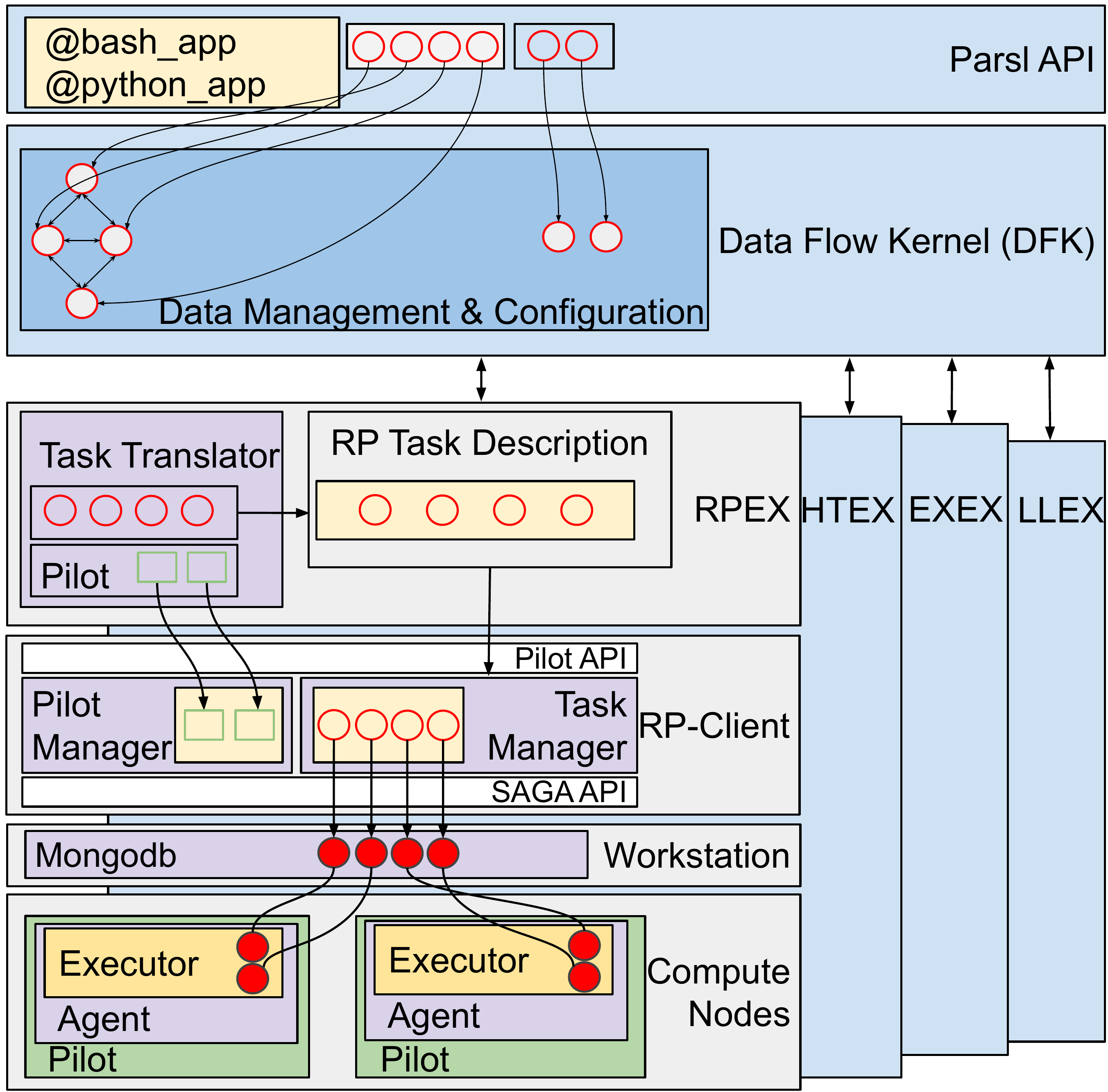}
    \caption{RPEX integration architecture.}
    \label{fig:int-arch}
    \up
\end{figure}

We implement a new Parsl executor for RP shown in Fig.~\ref{fig:int-arch} and we
call it RADICAL-Pilot Executor (RPEX).
RPEX is a Python class that bootstraps RP when initialized by Parsl. To make
RPEX consistent with other Parsl executors, we based RPEX's implementation on
the Parsl HTEX executor class.

\kyle{Ah, this is where I was confued above. I wonder if this should go up a
section into the discussion of how the models are different}\mtnote{we run out
of time. Meanwhile, I moved it here.} Note that Parsl does not require resource
specification at task level, while RP requires specification of the number of
cores and threads for both CPU cores and GPUs for every task.
To enable the use of RP's resource management capabilities in RPEX, we extended
Parsl's API to allow users to define those parameters. \kyle{Great, could
mention what that looks like as a nice contribution to Parsl, and potentialyl
relate to how its done with WQEX if at all compatible} \logan{I made a Parsl
issue about documenting this effort}\aanote{We will work on documenting it}

Once Parsl starts, it initializes DFK and RPEX simultaneously. Upon
initialization, the DFK: (i) obtains the tasks from Parsl API; (ii) builds the
tasks table; (iii) solves each task dependencies; and (iv) passes each task
object to RPEX for execution (Fig.~\ref{fig:int-arch}). Once initialized, RPEX:
(i) obtains via its interface the HPC platform on which to execute the tasks and
the amount of walltime for which to hold the resources; (ii) starts a new RP
session and creates the Pilot Manager and the Task Manager; and (iii) obtains
the number of CPU cores and GPUs required by each task
submitted by the DFK.

Submitting and executing Parsl tasks via RP require translating those tasks into
task objects that can be interpreted by RP. Each task object has a set of
properties, e.g., the executable's name, its type, its arguments (if any), the
number and type of resources, the number of processes, etc. Once the DFK submits
the Parsl task to RPEX, the Task translator unpacks, translates and maps that task
to the corresponding RP task object.

Parsl's DFK monitors the status of RPEX and, once ready, it starts submitting
Parsl tasks one by one to RP (Fig.\ref{fig:int-arch}, RP-Client). Eventually, RP
submits the task to its Task Manager to be scheduled and executed on the pilot
resource (Fig.\ref{fig:int-arch}, Compute Nodes).


\subsection{RP MPI Function Executor}

\kyle{Unclear what the executor is here, I think it is RP-specific and thus not
implementing the executor interface we have talked about until now. Any chance
it could be renamed?}\aanote{this is RP specific executor.}

\begin{figure}[h]
    \centering
    \includegraphics[width=0.33\textwidth]{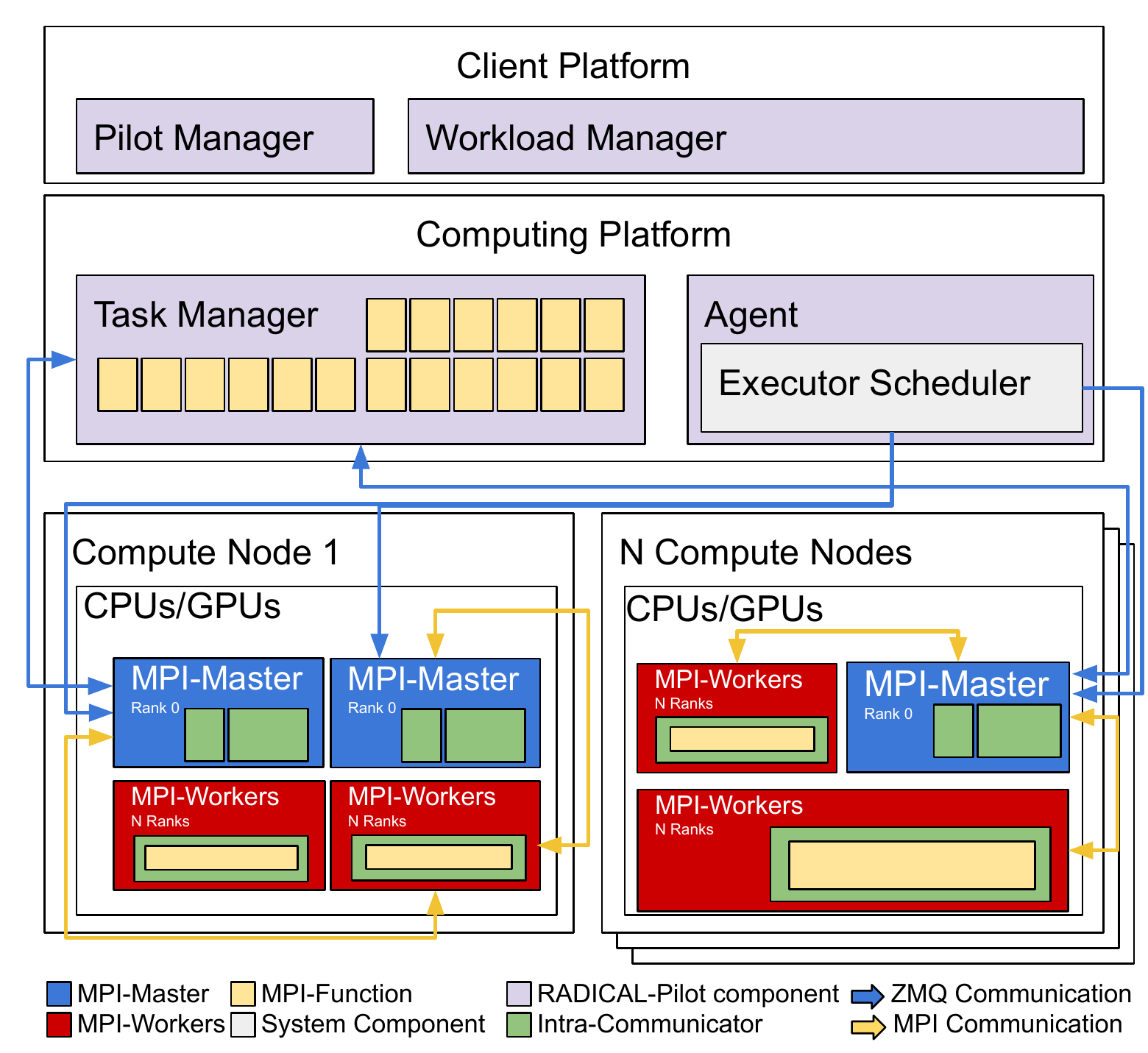}
    \caption{RADICAL-Pilot multi-node MPI function executor.}
    \label{fig:rp-exec-mpi}
    \up
\end{figure}

To support launching and executing of MPI Python functions, we implemented a RP
single/multi-node MPI function executor shown in Fig.~\ref{fig:rp-exec-mpi}. Not
to be confused with RPEX, that is a Parsl executor, the RP executor uses
\texttt{mpi4py}
to implement a task-based SPMD master-worker paradigm to concurrently execute
heterogeneous MPI Python functions.

The MPI executor communicates with other RP's component via ZeroMQ, sending and
receiving MPI Python functions, until terminated by RP's Agent. Once
scheduled by RP, the MPI function executor:
launches itself via a user-specified MPI launch method; loads the
\texttt{mpi4py} environment once for all incoming tasks; and spawn the
MPI-Master and MPI-Workers.

The MPI-Master decomposes the main MPI-Communicator into several
Intra-Communicators to serve as a private communicator for every Python
function. Once the MPI-Master receives the functions via a ZMQ channel, it sends
them to the designated workers for execution. Workers can concurrently run on
single or multiple nodes. Every MPI-Master is responsible for coordinating the
execution of a set of Python functions and performing MPI collective
communications among the workers (see Fig.~\ref{fig:rp-exec-mpi}).

\kyle{The relation to the RPEX is also not clear hear. Is this something that
can be used via the Parsl-RP integration?}\aanote{Yes, this is something we did
use in one of the usecases via RP-Parsl}


\mtnote{I propose to comment out the following: ``Once the workers starts
executing the MPI Python functions, the MPI-Master mark these workers as
\texttt{BUSY}. Similarly, once a worker finishes a task execution, the master
marks that worker as \texttt{IDLE} to be reused by other workers.''}\aanote{fine by me}



\section{Experiments and Evaluation}
\label{sec:experiments}

Table~\ref{tab:exp_table} shows the setup of our experiments. We use both SDSC
Expanse and TACC Frontera for Experiment 1, and Frontera only for Experiment 2.
Expanse compute nodes have 128 CPU cores, while Frontera has two types of nodes:
``normal'' with 56 CPU cores and no GPUs, and ``rtx'' with 16 CPU cores and 4
GPUs.

\begin{table*}
	\up
	\caption{Setup for experiments 1 and 2. WS/SS = weak/strong scaling;
	COL = Colmena; IWP = Ice Wedge Polygons.}
	\label{tab:exp_table}
	\centering
	\begin{tabular}{lllclcc}
		\toprule
		ID                                  &
		Experiment Type                     &
		Platform                            &
		Nodes                               &
		Task Type                           &
		CPUs(cores)                         &
		GPUs                                \\
        \midrule
		\multirow{2}{*}{1}                  &
        \multirow{2}{*}{MPI executor WS/SS} &
        Expanse                             &
        $2^1-2^5$                           &
        \multirow{2}{*}{MPI-Homogeneous}    &
        \#nodes $\times$ 128                    &
        \multirow{2}{*}{N/A}                \\
		                                    &
                                            &
        Frontera                            &
        $2^2-2^9$                           &
                                            &
        \#nodes $\times$ 56                     &
                                            \\
        \midrule
		2                                   &
        COL/IWP  WS/SS                      &
        Frontera                            &
        $2^5-2^8$                           &
        MPI-Heterogeneous                   &
        \#nodes $\times$ 56                     &
        $2^3-2^5$                           \\
		\bottomrule
	\end{tabular}
	\UP
\end{table*}

We use three metrics: Total Processing Time (TPT); Throughput (TS); and Total
Time to Execution (TTX). TPT is the time spent by our executor to finish
executing all the tasks of a workload. TS is the number of tasks executed per
second, calculated by dividing the total number of tasks by TPT. TTX is the
total amount of time taken by all tasks to finish executing. Note that TPT
measures the time in which the executor kept the resources busy, excluding any
idle or wait time. In contrast, TTX measures the time the workload spent to
finish the execution of all tasks on those resources, including idle and wait
time.

Experiment 1 measures the TPT and TS of the MPI-function executor presented
in~\S\ref{sec:integration}, as a function of the number of tasks. Experiment 2
measures TTX and RPEX integration overheads with the two use cases described
in~\S\ref{sec:usecases}. Together, these experiments enable us to characterize
the performance of the integrated RPEX and of the Python MPI function
executor with different resources, task heterogeneity, and scale on HPC
resources.

\subsection{Experiment 1: MPI-Function Executor Scalability}

We characterize the performance of the MPI-function executor on Expanse and
Frontera, measuring its strong and weak scaling in terms of TPT and TS. We
summarize the results in Table~\ref{tab:exp3_table}.\kyle{Reading scaling
results from a table is sort of difficult, maybe better to point at the graphs
first? and use the table for throughput?} Note that the processing time of the
MPI function executor measured by TPT includes the aggregated overheads of
launching the MPI infrastructure and of the MPI communications.

We use a homogeneous workload of Python \texttt{no-op} functions. We launch the
executor with \texttt{mpirun} and configure it to execute each MPI function
across two compute nodes, using \texttt{mpi4py} for each function. Every
function is configured to run on 256 ranks on Expanse (128 cores per node) and
112 ranks on Frontera (56 cores per node).  In this way, each task tests our
executor's multi-node capability, scaling to sizable portions of the HPC
platforms.

\begin{table}
	\up
	\centering
	\caption{Experiment 1 strong and weak scaling results on Expanse and Frontera.
		N = number of compute nodes.}
	\label{tab:exp3_table}
	\begin{tabular}{llcr @{\hspace{1\tabcolsep}} lr @{\hspace{1\tabcolsep}} l}
		\toprule
		                   &
		               &
		                     &
		\multicolumn{2}{c}{Total processing}                     &
	    \multicolumn{2}{c}{Throughput }                     \\
		System                   &
		Scaling               &
		N                     &
		\multicolumn{2}{c}{time (seconds)}                     &
	    \multicolumn{2}{c}{(tasks/second)}                      \\
		\midrule
		\multirow{8}{*}{Expanse} &
		\multirow{4}{*}{Strong} &
		2                       &
		6752.4 & $\pm153.9$        &
		4.7 & $\pm0.1$             \\
		&
		                       &
		4                       &
		3494.4 & $\pm199.0$          &
		9.2 & $\pm0.4$             \\
		&
		                        &
		8                       &
		1758.4 & $\pm88.5$         &
		18.3 & $\pm0.8$            \\
		                        & &
		16                      &
		911.3 & $\pm43.0$            &
		35.3 & $\pm1.6$            \\

		\cmidrule{3-7}
		&
		\multirow{4}{*}{Weak} &
		2                       &
		409.5 & $\pm4.9$           &
		4.8 & $\pm0.05$            \\
		&
		                        &
		4                       &
		423.1 & $\pm9.4$           &
		9.4 & $\pm0.2$             \\
		&
		                      &
		8                     &
		412.1 & $\pm2.5$         &
		19.4 & $\pm0.1$          \\

		&
		&
		16                    &
		430.5 & $\pm4.1$         &
		37.1 & $\pm0.3$          \\
	    &
	    &
		32                     &
		423.5 & $\pm4.8$        &
		75.5 & $\pm0.8$         \\
		\midrule
		%
		\multirow{12}{*}{Frontera} &
		\multirow{6}{*}{Strong} &
		8                         &
		14173.1 & $\pm375.2$         &
		36.1 & $\pm0.9$              \\
		&
		&
		16                      &
		7458.4 & $\pm341.9$        &
		69.0 & $\pm2.8$              \\
		&
		&
		32                      &
		3546.8 & $\pm105.6$        &
		144.7 & $\pm4.0$            \\

	         &
		&
		64                      &
		2035.3 & $\pm97.8$         &
		235.2 & $\pm11.5$          \\

		&&
		128                     &
		1236.8 & $\pm150.6$        &
		431.6 & $\pm51.4$          \\

		&&
		256                    &
		509.1 & $\pm8.6$          &
		1005.8 & $\pm17.1$        \\
		\cmidrule{3-7}
		&
		\multirow{7}{*}{Weak} &
		8                         &
		231.3 & $\pm6.1$             &
		34.6 & $\pm0.8$              \\
		&
		&
		16                       &
		228.8 & $\pm5.2$            &
		70.0 & $\pm1.6$               \\
		&
		&
		32                      &
		221.9 & $\pm4.4$           &
		144.2 & $\pm2.8$           \\
		&&
		64                     &
		238.5 & $\pm14.0$           &
		270.0 & $\pm15.6$           \\
		&&
		128                     &
		258.3 & $\pm14.5$         &
		498.3 & $\pm26.7$          \\

		&&
		256                     &
		309.4 & $\pm50.3$          &
		868.9 & $\pm129.3$         \\

		&&
		512                     &
		303.7 & $\pm17.5$          &
		1696.7 & $\pm94.2$         \\
		\bottomrule
	\end{tabular}
	\UP
\end{table}

\subsubsection{SDSC Expanse}\label{sssec:exp3_expanse}

\begin{figure*}[t]
	\centering
	\includegraphics[width=1\textwidth]{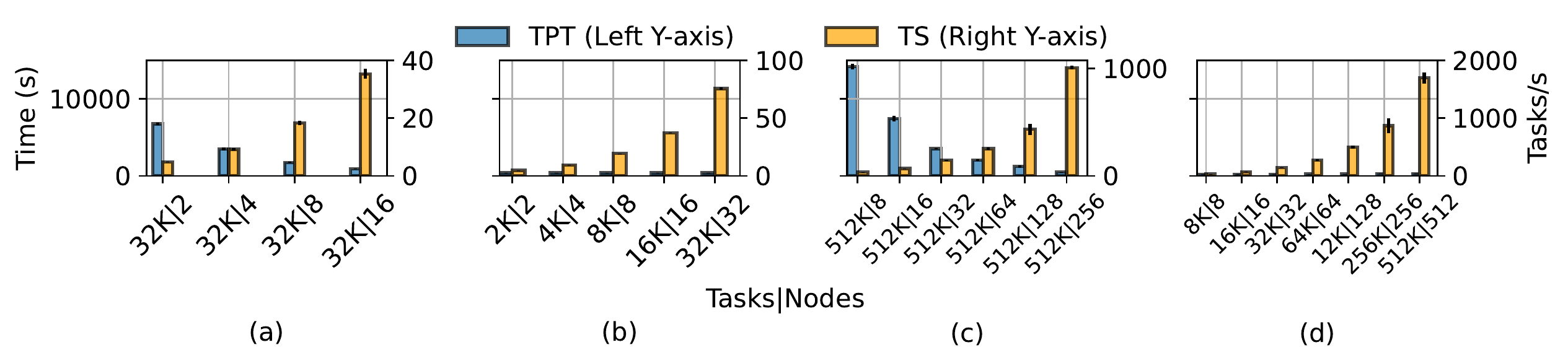}
	\caption{Experiment 1: Scaling properties of MPI Function executor. (a) and
	(b) characterize strong and weak scaling on Expanse, respectively; (c) and
	(d) characterize strong and weak scaling on Frontera, respectively.
	\kyle{Need to explain what the x axis labels mean} \kyle{I wonder if it
	would help to have an ideal scaling line so we can see how close performance
	is}}
	\label{fig:exp3_tpt_thp_frontera_expanse}
\end{figure*}

Fig.~\ref{fig:exp3_tpt_thp_frontera_expanse} shows the strong (a) and the weak
(b) scaling of our MPI executor in terms of TPT (blue) and TS (orange).
For strong scaling, TPT decreases linearly while TS increases linearly as a
function of the number of nodes across all runs (see
Table~\ref{tab:exp3_table}). Both TPT and TS show a consistent behavior across
the experiment's scale while maintaining small error bars, indicating an
efficient scaling behavior.


In the weak scaling case, the TPT shows consistent scaling with small error bars
across all runs. Further, the TS increases linearly with the number of nodes,
reaching a maximum of 75.5 task/s on 32 nodes. The linear increase of TS in
Fig.~\ref{fig:exp3_tpt_thp_frontera_expanse} (a,b) show a positive correlation
between the number of nodes and tasks, indicating that the executor can achieve
higher TS at larger scales.


Fig.~\ref{fig:exp3_tpt_thp_frontera_expanse} (a,b) show that our executor
scales efficiently on the assigned resource, using a homogeneous workload in
which each function requires the same number of ranks. We use a homogeneous
workload as it allows us to study the baseline performance of the executor since
this type of workload is more generalizable and easier to measure across scales.

\subsubsection{TACC Frontera}

We used the same experiment setting of Expanse to characterize strong and weak
scaling of the MPI function executor on Frontera at larger scale (see
Table~\ref{tab:exp3_table}). Fig.~\ref{fig:exp3_tpt_thp_frontera_expanse} (c,d)
show the strong and the weak scaling of the MPI executor's TPT (blue) and TS
(orange).



In the strong scaling case, TPT decreases linearly and TS increases
exponentially with the number of nodes, with small error bars across all runs.
In the weak scaling case, TPT remains stable from 8 to 64 nodes and then
increases sub-linearly from 128 to 512 nodes. Note that the error bars of 8--64
nodes overlap with the ones of 128--512 nodes, making the difference between
TPTs statistically less significant. The sub-linear increase in TPT between 128
and 512 nodes is due to the MPI collective communication overheads, leading to
TPT deterioration with increasing numbers of
resources~\cite{hfast2018}.
TS shows an exponential increase between 8 and 512 nodes while maintaining
relativity small error bars across all runs. This confirms our analysis
in~\S\ref{sssec:exp3_expanse} which shows that our executor reaches higher TS on
larger scales consistently.




Comparing results between Expanse and Frontera
we see that (i) as expected,
the cost of constructing and launching an MPI task grows with the number of
ranks (256 ranks vs.\ 112 ranks), since MPI takes more time to group and
construct a larger MPI-communicator~\cite{dinan2012an}; (ii) TPT increases
proportionally with the number of resources due to the increased number of MPI
communication overheads, and the number of ranks per task.



The MPI executor scales efficiently between 8 and 64 nodes on Frontera and 2 and
32 nodes on Expanse, but given the proportional increase of the overheads
between 128 and 512 nodes, performance starts to deteriorate at a larger scale
on Frontera (see Table~\ref{tab:exp3_table}). Constructing an MPI
Intra-communicator for every function is expensive but necessary in presence of
heterogeneous functions. The impact of that overhead on the overall workload
execution depends on the duration of the functions execution. For short-running
homogeneous functions, a more performant design would: (i) set up the
Intra-communicator only once to be reused by every task; (ii) take advantage of
caching capabilities for MPI Intra- and Inter-communicator.

\subsection{Experiment 2: Use Case Scalability}\label{ssec:exp4}

Next we study RPEX strong and weak scaling by running the Colmena and IWP use
cases (\S\ref{sec:usecases}) on Frontera while varying both problem size and
number of nodes.

Both use cases require capabilities that can be provided only by RPEX and not by
either RP or Parsl alone. Colmena requires execution of a workflow of concurrent
heterogeneous single-node MPI executables and single core non-MPI Python
functions. IWP requires instead a workflow of heterogeneous MPI Python functions
that run concurrently on multiple nodes.

We summarize our experiment results in Table~\ref{tab:exp4_table},
using three metrics: RP overheads, RPEX integration overheads, and total time to
completion (TTX). Note that we measure RPEX
overhead as the sum of Parsl and RP overheads. Parsl's overhead includes the
amount of time taken to: (1) start the executor; (2) build the DAG of tasks; (3)
solve the data dependencies among all tasks; (4) submit the tasks to the
executor; and (5) shutdown and cleanup both the executor and the integration
components. RP's overhead consists of the amount of time taken to: (1) start the
runtime system; and (2) manage the tasks' execution.

\begin{table}
\setlength{\tabcolsep}{4pt}
	\up
	\centering
	\caption{Experiment 2 strong and weak scaling results on Frontera.
	N = number of compute nodes. Times
	are in seconds.}
	\label{tab:exp4_table}
	\begin{tabular}{llcr @{\hspace{1\tabcolsep}} lr @{\hspace{1\tabcolsep}} lr @{\hspace{1\tabcolsep}} l}
		\toprule
		               &
		 &
		                    &
		\multicolumn{2}{c}{Total time}                     &
		\multicolumn{2}{c}{RP}                 &
		\multicolumn{2}{c}{RPEX}            \\
		App               &
		Scaling	  &
		N                     &
		\multicolumn{2}{c}{to execution}                    &
		\multicolumn{2}{c}{overhead}                  &
		\multicolumn{2}{c}{overhead}            \\

		\midrule
		%
		\multirow{8}{*}{\rot{Colmena}} &
		\multirow{3.5}{*}{Strong} &
		32                      &
		8725.6 & $\pm$7.0            &
		104.4 & $\pm$4.5           &
		724 & $\pm$7.5            \\
		%
								&&
		64                      &
		3961.9 & $ \pm$2.0            &
		116.4 & $ \pm$11.8          &
		744.6 & $ \pm$13.5          \\
		%
		                        &&
        128                     &
		1929.2 & $ \pm$19.0           &
		99.7 & $ \pm$5.0              &
		769.1 & $ \pm$10.5          \\

		                        &&
		256                     &
		3263.9 & $ \pm$94.7         &
        		176.2 & $ \pm$56.4          &
        		855.2 & $ \pm$56.2          \\
		\cmidrule{3-9}
		    &
		\multirow{3.5}{*}{Weak} &
		32                      &
		620.9 & $ \pm$1.3          &
		173.1 & $ \pm$0.2          &
		217.5 & $ \pm$2.1          \\
		%
                                &&
		64                      &
		629.1 & $ \pm$0.9           &
		118.8 & $ \pm$1.4           &
		207.8 & $ \pm$1.7           \\
		%
		                        &&
		128                     &
		636.5 & $ \pm$0.6           &
		233.7 & $ \pm$34.7          &
		253.0 & $ \pm$37.1           \\

		                        &&
		256                     &
		1891.8 & $ \pm$61.6         &
		134.8 & $ \pm$2.0             &
		388.2 & $\pm$74.3          \\
		\midrule
		\multirow{8}{*}{\rot{Ice Wedge Polygons}} &
		\multirow{3.5}{*}{Strong} &
		2                       &
		10620.3 & $\pm$34.0        &
		6.3 & $\pm$2.1           &
		7.1 & $\pm$2.1          \\
		                        &&
		4                       &
		4895.9 & $\pm$1.4         &
		5.5 & $\pm$3.0               &
		6.4 & $\pm$2.1           \\
		                        &&
		8                       &
		2460.9 & $\pm$4.9         &
		4.6 & $\pm$2.3           &
		5.4 & $\pm$2.4            \\
		                        &&
		16                      &
		1344.6 & $\pm$51          &
		5.5 & $\pm$2.3           &
		6.4 & $\pm$0.1            \\
       \cmidrule{3-9}
		&
		\multirow{3.5}{*}{Weak} &
		2                       &
		211.6 & $\pm$5.6         &
		5.7 & $\pm$1.5            &
		5.7 & $\pm$1.6           \\
		                        &&
		4                       &
		236.1 & $\pm$0.1         &
		6.0 & $\pm$1.0           &
		6.0 & $\pm$1.1           \\
		                        &&
		8                       &
		259.9 & $\pm$1.9         &
		7.4 & $\pm$1.9           &
		7.4 & $\pm$1.9           \\

                                &&
        		16                      &
        		275.0 & $\pm$11.8         &
        		6.2 & $\pm$1.4           &
        		6.3 & $\pm$1.4           \\
        \bottomrule
	\end{tabular}
	\UP
\end{table}

\subsubsection{Colmena}

We created a heterogeneous synthetic workflow based on a real-world Colmena
application~\cite{ward_colmena_2021} to evaluate the new capabilities provided
by RPEX. The workflow consists of three tasks: Python ``pre-process'' and
``post-process'' functions, and a C MPI ``simulation'' executable that runs for
$\sim$100s. The pre-process function prepares the execution environment for the
simulation MPI task, while the post-processing function collects results from
the simulation tasks, storing them in a Python class object. Each pre-process
and post-process function requires one CPU core, while each simulation task
requires a full node (56 CPU cores). We used the TACC-specific MPI launcher
\texttt{Ibrun} to launch the MPI executables.

\begin{figure*}[t]
	\centering
	\includegraphics[width=1\textwidth]{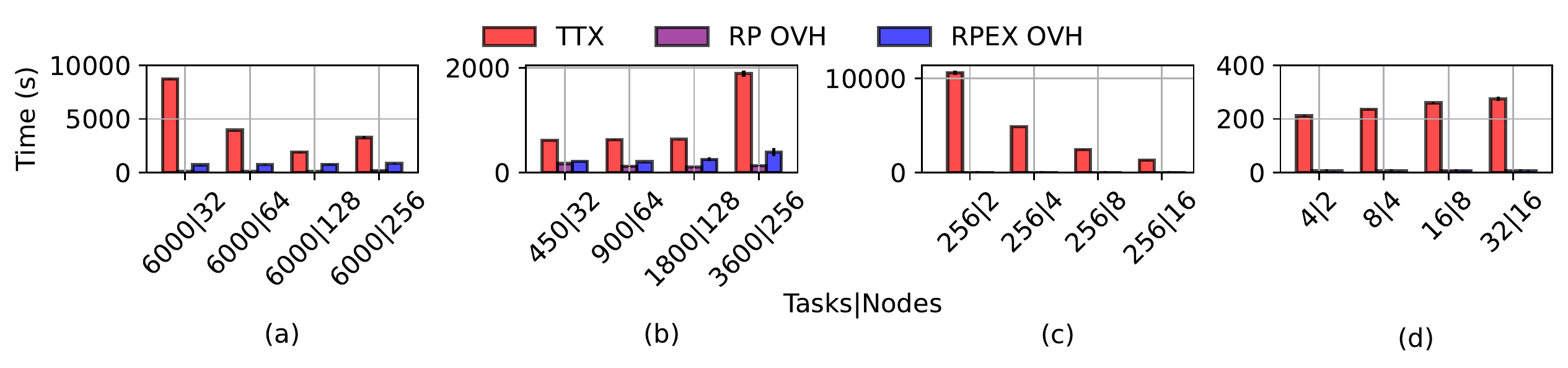}
	\UP
	\caption{Experiment 2: Scaling properties of Colmena and Ice Wedge Polygons
	(IWP). (a) and (b) characterize strong and weak scaling of Colmena; (c) and
	(d) characterize the strong and weak scaling in~\S\ref{sec:usecases}. RPEX
	overheads include RP overheads and represent the total overhead of the
	execution. RP overheads represent the time spent by RP and its executor to
	manage the execution of the workload/workflow.}\label{fig:exp4_col_iwp_ss_ws}
	\UP
\end{figure*}

Fig.~\ref{fig:exp4_col_iwp_ss_ws} shows strong (a) and weak (b) scaling with
RPEX executing the Colmena workflow. TTX (red) decreases linearly between 32 and
128 nodes in strong scaling and maintains a consistent scale in the weak
scaling.
RP overheads (purple) are essentially invariant of scale, while RPEX overheads
(blue) increase with scale in the weak scaling and maintains a relativity
consistent behavior in the strong scaling. The scaling behavior changes at 256
nodes for both strong and weak scaling, showing a linear increase of TTX
compared to the 32--128 nodes run. We investigated the TTX increase by measuring
the resource utilization of Colmena workflows while executing 450/32, 900/64,
1800/128 and, 3600/256 tasks/nodes.

\begin{figure*}[t]
	\centering
	\includegraphics[width=1\textwidth]{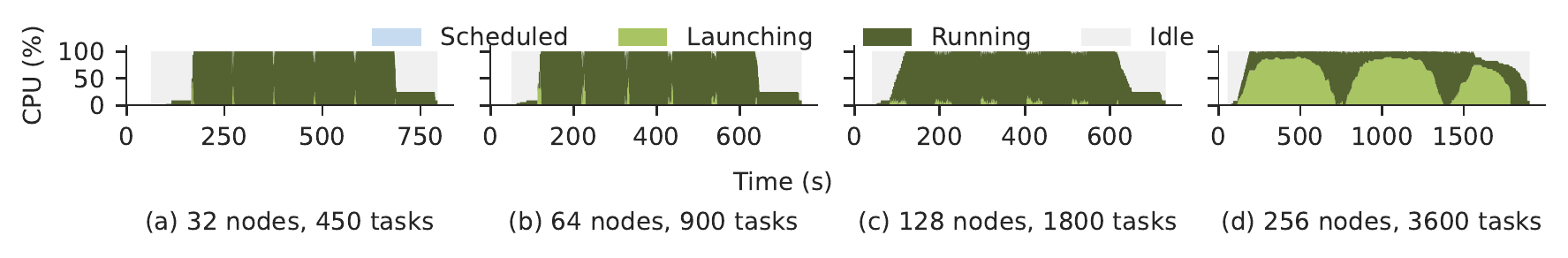}
	\UP
	\caption{Experiment 2: Colmena resource utilization with RPEX on 32, 64, 128
	and 256 nodes}\label{steady_state}
	\label{fig:colmena_res_utilz}
	\UP
\end{figure*}

In Fig.~\ref{fig:colmena_res_utilz}, we measure and break down the resource
utilization of Colmena's TTX based on four task-related events:
``Scheduled'' indicates resources being assigned to tasks and ready for
execution; ``Launching'' indicates the resources occupied while waiting for
\texttt{Ibrun} to launch the scheduled tasks by RP and also represents
\texttt{Ibrun} overheads; ``Running'', which shows the resources occupied by RP
while executing the launched tasks;
and ``Idle'' the time in which the available resources are occupied but not
busy.

Fig.~\ref{fig:colmena_res_utilz} (a,b,c) show that Running (dark green) occupies
$\sim$98\% of the available resources with average Launching time of 30s, 65s
and 215.3s. Fig.~\ref{fig:colmena_res_utilz} (d) shows instead that Launching
(light green) becomes the dominant activity, occupying most of the resources for
$\sim$1791.2s. Launching creates a `busy wait' condition that prevents tasks
from executing. This explains the increase in Colmena's TTX shown in
Fig.~\ref{fig:exp4_col_iwp_ss_ws} (a,b): RP takes longer to execute on 256
nodes compared to 32--128 nodes because it has to wait longer for \texttt{Ibrun}
to launch the tasks.

Comparing Colmena's resource utilization from prior
work~\cite{ward_colmena_2021}---while executing only Python functions without
RP---to Fig.~\ref{fig:colmena_res_utilz} shows comparable resource utilization.
RPEX reaches a resource utilization of $\sim99$\% while executing both
MPI-executables and Python functions, maintaining the performance measured when
Colmena executed only non-MPI functions via Parsl. RPEX results in
Fig.~\ref{fig:exp4_col_iwp_ss_ws} also show that RPEX has the potential to reach
large scales with low and constant overheads (RP overhead). Substituting
\texttt{Ibrun} with more performant MPI libraries has the potential to lower the
task launching overheads at
scale~\cite{merzky2021design,turilli2019characterizing}.

\subsubsection{Ice Wedge Polygons (IWP)}

We implemented the IWP use case of \S\ref{sec:usecases} by using the Single
Program Multiple Data (SPMD) MPI pattern, where tasks are split up and
concurrently executed on multiple cores with different
inputs~\cite{SPMD_Darema2011}. We configured RPEX to use RP's MPI function
executor, executing the IWP workload with the SPMD pattern. We used 2 GPUs and 8
CPU cores per task, with up to 256 concurrent tasks on Frontera.

Fig.~\ref{fig:exp4_col_iwp_ss_ws} shows the strong (c) and weak (d) scaling of
RPEX with the IWP use case workload. In the strong scaling case, TTX (red)
decreases exponentially. In the weak scaling case, TTX shows a sublinear
increase across all runs (see Table~\ref{tab:exp4_table}). As with the Colmena
use case, RP overheads shows a consistent behavior across all the runs. Further,
RPEX overheads in the strong and weak scaling maintains a consistent behavior
while executing on $\le$ 128 nodes. The scaling of the TTX confirms that
\texttt{Ibrun} overheads increases marginally with the number of tasks and
resources when executing on $\le$ 128 nodes and becomes intolerable with $>$ 128
node runs. Overall, RP and RPEX overheads are low compared to the TTX of the use
case and the scale of the experiments.




\section{Conclusions and Future Work}
\label{sec:conclusion}

We described an integration of RADICAL-Pilot's pilot capabilities with Parsl's
workflow management capabilities to enable the execution of production workflows
on diverse HPC platforms. The four main contributions of this paper are: (1) a
case study of the engineering process used to integrate two independently
developed middleware systems; (2) an analysis of the design of an executor for
MPI Python functions tailored to HPC resources; (3) a performance
characterization of both the integrated system and the proposed executor on two
HPC platforms; and (4) a description of how the integrated system and executor
have been used for two exemplar use cases.

The RPEX integration shows that integrating independent middleware systems
requires an analysis of their capabilities, execution and state models, private
and public APIs, and performance bottlenecks. We showed that, based on that
analysis, the integration's engineering effort can be reduced, while avoiding
expensive rewriting of existing code bases. Nonetheless, we also showed that
user-facing APIs might have to be extended to make specific information
available across the integrated systems. \kyle{I dont understand this final
sentence}\mtnote{we had to extend Parl API to allow expressing resource
requirements. I refrased, better?}

We successfully supported two classes of use case.
First, we used RPEX in Colmena, enabling efficient execution
of both MPI and single node Python functions.
Importantly, no changes were required to Colmena to make use of RPEX.
Second, by supporting IWP, we showed how RPEX can be used out of
the box to code specific workflows and run them at scale. 
Together, these results confirm that the integration of independent middleware
components can be a viable approach to reducing capability duplication across
middleware, especially when considering the challenges posed by executing
workflows in production on Exascale HPC platforms.

Our experiments show that RPEX overheads increase only marginally with the
number of tasks and resources. This result shows that our integration approach,
based on the weak coupling of the two systems via a light-weight, stand-alone
interface, does not introduce major overheads when executing heterogeneous
workflows on HPC resources. It also highlights highlights the limitations of the
MPI library used to launch tasks at scale and establishes the need for
developing workflow-specific low-level communication libraries.
\jhanote{(1) There is a repeat of highlight in the last two sentences. (2) I'm
not sure if we can merge the two sentences? If so, proposed sentence: ``..
highlights the limitations of the MPI library used to launch tasks at scale and
establishes the need for developing workflow-specific low-level communication
libraries.''}\mtnote{thanks. Done.}


The performance of RPEX can be improved by adopting a different task submission
logic. Currently, RPEX submits a stream of tasks to the runtime engine which
increases overhead at scale, especially with short running tasks. We are
developing a \texttt{bulk submission} mode for RPEX, which can reduce overheads,
improving task submission, scheduling, and execution throughput.\kyle{I dont
know what ``as a workload'' means?}\aanote{thanks, is this better?}\jhanote{I
think -- if at all possible, linking this work to be viewed as ``software
sustainability engineering'' would be good, but I know space is exceedingly
tight, thus just a suggestion.}

\vspace*{1em}
\noindent\footnotesize{This work was supported by the ECP ExaWorks and ExaLearn
projects, as well as NSF-1931512 (RADICAL-Cybertools). HPC access on XSEDE was
provided by allocation TG-MCB090174.}

\bibliographystyle{IEEEtran}
\bibliography{main}

\begin{thebibliography}{10}
\providecommand{\url}[1]{#1}
\csname url@samestyle\endcsname
\providecommand{\newblock}{\relax}
\providecommand{\bibinfo}[2]{#2}
\providecommand{\BIBentrySTDinterwordspacing}{\spaceskip=0pt\relax}
\providecommand{\BIBentryALTinterwordstretchfactor}{4}
\providecommand{\BIBentryALTinterwordspacing}{\spaceskip=\fontdimen2\font plus
\BIBentryALTinterwordstretchfactor\fontdimen3\font minus
  \fontdimen4\font\relax}
\providecommand{\BIBforeignlanguage}[2]{{%
\expandafter\ifx\csname l@#1\endcsname\relax
\typeout{** WARNING: IEEEtran.bst: No hyphenation pattern has been}%
\typeout{** loaded for the language `#1'. Using the pattern for}%
\typeout{** the default language instead.}%
\else
\language=\csname l@#1\endcsname
\fi
#2}}
\providecommand{\BIBdecl}{\relax}
\BIBdecl

\bibitem{workflow-systems-url}
M.~R. Crusoe \emph{et~al.}, ``Computational data analysis workflow systems,''
  2022, \url{https://s.apache.org/existing-workflow-systems}.

\bibitem{workflows2021silva-1}
\BIBentryALTinterwordspacing
R.~Ferreira~da Silva, H.~Casanova \emph{et~al.},
  ``\BIBforeignlanguage{en}{Workflows community summit: Advancing the
  state-of-the-art of scientific workflows management systems research and
  development},'' Tech. Rep., 2021. [Online]. Available:
  \url{https://zenodo.org/record/4915801}
\BIBentrySTDinterwordspacing

\bibitem{babuji2019parsl}
Y.~Babuji, A.~Woodard \emph{et~al.}, ``Parsl: Pervasive parallel programming in
  {P}ython,'' in \emph{28th International Symposium on High-Performance
  Parallel and Distributed Computing}.\hskip 1em plus 0.5em minus 0.4em\relax
  ACM, 2019, p. 25–36.

\bibitem{merzky2021design}
A.~Merzky, M.~Turilli \emph{et~al.}, ``Design and performance characterization
  of {RADICAL-Pilot} on leadership-class platforms,'' \emph{IEEE Transactions
  on Parallel and Distributed Systems}, vol.~33, no.~4, pp. 818--829, 2021.

\bibitem{ward_colmena_2021}
L.~Ward, G.~Sivaraman \emph{et~al.}, ``Colmena: Scalable machine-learning-based
  steering of ensemble simulations for high performance computing,'' in
  \emph{IEEE/ACM Workshop on Machine Learning in High Performance Computing
  Environments}, 2021, pp. 9--20.

\bibitem{turilli2019middleware}
M.~Turilli, V.~Balasubramanian \emph{et~al.}, ``Middleware building blocks for
  workflow systems,'' \emph{Computing in Science \& Engineering}, vol.~21,
  no.~4, pp. 62--75, 2019.

\bibitem{wang2009kepelr_hadoop}
J.~Wang, D.~Crawl, and I.~Altintas, ``Kepler + {H}adoop: A general architecture
  facilitating data-intensive applications in scientific workflow systems,'' in
  \emph{4th Workshop on Workflows in Support of Large-Scale Science}.\hskip 1em
  plus 0.5em minus 0.4em\relax ACM, 2009.

\bibitem{Luckow_2016_Hadoop}
A.~{Luckow}, I.~{Paraskevakos} \emph{et~al.}, ``Hadoop on {HPC}: Integrating
  {H}adoop and pilot-based dynamic resource management,'' in \emph{IEEE
  International Parallel and Distributed Processing Symposium Workshops}, 2016,
  pp. 1607--1616.

\bibitem{badia2022pycompss}
R.~M. Badia, J.~Conejero \emph{et~al.}, ``Pycompss as an instrument for
  translational computer science,'' \emph{Computing in Science \& Engineering},
  vol.~24, no.~2, pp. 79--84, 2022.

\bibitem{wilde2011swift}
M.~Wilde, M.~Hategan \emph{et~al.}, ``Swift: A language for distributed
  parallel scripting,'' \emph{Parallel Computing}, vol.~37, no.~9, pp.
  633--652, 2011.

\bibitem{turilli2017evaluating}
M.~Turilli, Y.~N. Babuji \emph{et~al.}, ``Evaluating distributed execution of
  workloads,'' in \emph{13th International Conference on e-Science}.\hskip 1em
  plus 0.5em minus 0.4em\relax IEEE, 2017, pp. 276--285.

\bibitem{shaffer2021lightweight}
T.~Shaffer, Z.~Li \emph{et~al.}, ``Lightweight function monitors for
  fine-grained management in large scale {P}ython applications,'' in \emph{Int.
  Parallel and Distributed Processing Symposium}.\hskip 1em plus 0.5em minus
  0.4em\relax IEEE, 2021, pp. 786--796.

\bibitem{bui2011work}
P.~Bui, D.~Rajan \emph{et~al.}, ``Work {Q}ueue + {P}ython: A framework for
  scalable scientific ensemble applications,'' in \emph{Workshop on Python for
  High Performance and Scientific Computing at SC11}.\hskip 1em plus 0.5em
  minus 0.4em\relax Citeseer, 2011.

\bibitem{peterson2019merlin}
J.~L. Peterson, R.~Anirudh \emph{et~al.}, ``Merlin: Enabling machine
  learning-ready {HPC} ensembles,'' 2019, arXiv 1912.02892.

\bibitem{Maestro_2019}
F.~D. Natale, ``Maestro workflow conductor,''
  \url{https://github.com/LLNL/maestrowf}, 2019.

\bibitem{Dong_2014_Flux}
D.~H. {Ahn}, J.~{Garlick} \emph{et~al.}, ``Flux: A next-generation resource
  management framework for large {HPC} centers,'' in \emph{43rd International
  Conference on Parallel Processing Workshops}, 2014, pp. 9--17.

\bibitem{guo2022composition}
J.~Guo, L.~Ward \emph{et~al.}, ``Composition-transferable machine learning
  potential for {LiCl-KCl} molten salts validated by high-energy x-ray
  diffraction,'' \emph{Physical Review B}, vol. 106, no.~1, p. 014209, 2022.

\bibitem{Witharana_2021_IWP}
C.~Witharana, M.~A.~E. Bhuiyan, and A.~K. Liljedahl, ``An object-based approach
  for mapping tundra ice-wedge polygon troughs from very high spatial
  resolution optical satellite imagery,'' \emph{Remote Sensing}, vol.~13,
  no.~4, 2021.

\bibitem{Turilli_2018_Comprehensive}
M.~Turilli, M.~Santcroos, and S.~Jha, ``A comprehensive perspective on
  pilot-job systems,'' \emph{ACM Computing Surveys}, vol.~51, no.~2, 2018.

\bibitem{concc_2020}
``Python 3.9.4 documentation: Concurrent.futures - launching parallel tasks,''
  2020.

\bibitem{Dalcn_2005_mpi4py}
L.~Dalcín, R.~Paz, and M.~Storti, ``{MPI for Python},'' \emph{Journal of
  Parallel and Distributed Computing}, vol.~65, no.~9, pp. 1108--1115, 2005.

\bibitem{turilli2016integrating}
M.~Turilli, F.~Liu \emph{et~al.}, ``Integrating abstractions to enhance the
  execution of distributed applications,'' in \emph{IEEE International Parallel
  and Distributed Processing Symposium}.\hskip 1em plus 0.5em minus 0.4em\relax
  IEEE, 2016, pp. 953--962.

\bibitem{hfast2018}
S.~H{\"{o}}finger, T.~Ruh, and E.~J. Haunschmid, ``Fast approximate evaluation
  of parallel overhead from a minimal set of measured execution times,''
  \emph{Parallel Processing Letters}, vol.~28, no.~1, pp.
  1\,850\,003:1--1\,850\,003:12, 2018.

\bibitem{dinan2012an}
J.~Dinan, D.~Goodell, and W.~Gropp, ``Efficient multithreaded context {ID}
  allocation in {MPI},'' in \emph{19th European Conference on Recent Advances
  in the Message Passing Interface}.\hskip 1em plus 0.5em minus 0.4em\relax
  Springer-Verlag, 2012, p. 57–66.

\bibitem{turilli2019characterizing}
M.~Turilli, A.~Merzky \emph{et~al.}, ``Characterizing the performance of
  executing many-tasks on {S}ummit,'' in \emph{IEEE/ACM Third Annual Workshop
  on Emerging Parallel and Distributed Runtime Systems and Middleware}.\hskip
  1em plus 0.5em minus 0.4em\relax IEEE, 2019, pp. 18--25.

\bibitem{SPMD_Darema2011}
F.~Darema, ``{SPMD} computational model,'' in \emph{Encyclopedia of Parallel
  Computing}.\hskip 1em plus 0.5em minus 0.4em\relax Springer US, 2011, pp.
  1933--1943.

\end{thebibliography}

\end{document}